# Towards optimal thermal distribution in magnetic hyperthermia

R. A. Rytov[1], V. A. Bautin[1] & N. A. Usov[1,2]

[1]*National University of Science and Technology «MISiS», 119049, Moscow, Russia*
[2]*Pushkov Institute of Terrestrial Magnetism, Ionosphere and Radio Wave Propagation, Russian Academy of Sciences, IZMIRAN, 142190, Troitsk, Moscow, Russia*

**Abstract**

A linear combination of spherically symmetric heat sources is shown to provide optimal stationary thermal distribution in magnetic hyperthermia. Furthermore, such spatial location of heat sources produces suitable temperature distribution in biological medium even for assemblies of magnetic nanoparticles with a moderate value of specific absorption rate (SAR), of the order of 100 - 150 W/g. We also demonstrate the advantage of using assemblies of magnetic nanocapsules consisting of metallic iron nanoparticles covered with non magnetic shells of sufficient thickness in magnetic hyperthermia. Based on numerical simulation we optimize the size and geometric structure of biocompatible capsules in order to minimize the influence of strong magneto-dipole interaction between closely spaced nanoparticles. It is shown that assembly of capsules can provide sufficiently high SAR values of the order of 250 - 400 W/g at moderate amplitudes $H_0$·= 50 - 100 Oe and frequencies $f$ = 100 - 200 kHz of alternating magnetic field, being appropriate for application in clinics.

## 1. Introduction

Magnetic hyperthermia is a new promising tool for cancer treatment [1-6]. In this method a proper amount of magnetic fluid containing superparamagnetic nanoparticles is introduced into a tumor. Under the influence of an alternating (ac) magnetic field, magnetic nanoparticles are able to locally heat a tumor to a temperature of the order of 42–45 °C required for medical indications. Laboratory and clinical studies show [7-11] that after several sessions of such procedure, in most cases the tumor stops growing and decays. An advantage of magnetic hyperthermia is the ability to affect tumors located deep in a human body, which is not possible for existing laser technologies [12,13]. In addition, magnetic hyperthermia can help to treat tumors with a poorly developed blood network, when the use of chemotherapy is ineffective.

For successful planning of therapeutic effects in magnetic hyperthermia, it is imperative to obtain a preliminary estimation of the stationary temperature distribution within a tumor and its vicinity. Proper planning of the heat sources power and their spatial distribution should ensure that the temperature in the entire tumor volume is within the therapeutic window 42 - 45°C, the overheating of healthy tissues surrounding the tumor being avoided. To calculate the stationary temperature distribution in an isotropic biological medium a spherically symmetric solution of the heat equation [14 - 17] is usually used. This approach was first applied by Andrä et al. [14] to study the temperature distribution during magnetic hyperthermia of breast cancer. In [15] a generalization of the spherical thermal model to the case of the presence of several spherical layers with different thermo-physical parameters was considered. In addition, fundamental solutions of the heat equation in rectangular, spherical, and cylindrical coordinate systems were obtained in [15,16]. In more complicated cases a solution of the Pennes bioheat equation [18 - 22] was studied. The latter approximately takes into account heat transfer from the heated area by blood flow.

It is worth mentioning that although the spherical thermal model [14-17] is attractive because of its simplicity, geometrically it is quite restrictive due to the small number of adjustable parameters. This makes it difficult to use spherical model to find optimal stationary thermal field in magnetic hyperthermia. On the other hand, detailed calculations based on the Pennes equation [18-22] require computer resources and time, and at the same time do not guarantee high accuracy due to the approximate nature of the Pennes equation itself. In this paper, it is shown that at least preliminary results for optimal stationary thermal distribution in magnetic hyperthermia can be obtained using a linear combination of spherically symmetric heat sources. It is shown that optimal spatial location of spherically symmetric heat sources produces necessary temperature distribution in biological medium even for assemblies of magnetic nanoparticles with a moderate value of specific absorption rate (SAR) of the order of 100 - 150 W/g.

Another important problem in magnetic hyperthermia is maintaining of sufficient thermal power of heat sources distributed in the biological media. Despite the large SAR values obtained for some assemblies of nanoparticles in a viscous liquid [23-25], it is not easy to ensure sufficient thermal power of these assemblies in biological medium. It is found [26-28] that magnetic nanoparticles form usually dense clusters inside biological cells and in intercellular space. This leads to a significant decrease in SAR due to strong magneto-dipole (MD) interaction between the nanoparticles [29-31]. In addition, under the influence of an aggressive medium weakly protected magnetic nanoparticles collapse rather quickly [4,32]. As a result, a significant decrease in the thermal power of the assembly may occur over time. Finally, there is a risk that a certain fraction of the magnetic nanoparticles introduced into a tumor can be removed from it by blood flow.

To overcome the above problems, in a number of works [33 - 36] it is proposed to use as heat sources in magnetic hyperthermia small capsules of biocompatible material, containing the optimal amount of magnetic nanoparticles. However, the size of capsules, as well as the magnetic and geometrical parameters of nanoparticles should be selected in such a way to minimize the reduction of the SAR due to particle MD interaction and to ensure the functioning of nanoparticles in the tumor with a given SAR value for several sessions of hyperthermia treatment. Using numerical simulation, in this paper we optimize the size and geometric structure of biocompatible capsules containing spherical iron nanoparticles in order to obtain sufficiently high SAR values, of the order of 250 – 400 W/g, at moderate amplitudes $H_0$ = 50 - 100 Oe and frequencies $f$ = 100 - 200 kHz of ac magnetic field most suitable [10,11] for application in clinics.

## 2. Results and discussion

**A. Stationary temperature distribution in a tumor**

Consider a homogeneous biological medium with an average thermal conductivity $k_s$, where permanent heat sources with a density $n(x,y,z)$ are distributed. A stationary temperature distribution in the medium is described by the heat equation

$$\left(\frac{\partial^2 T}{\partial x^2} + \frac{\partial^2 T}{\partial y^2} + \frac{\partial^2 T}{\partial z^2}\right) + \frac{1}{k_s} n(x,y,z) = 0 \quad (1)$$

As a boundary condition to this equation, one can assume a constant temperature on a certain surface $S$ located far enough from the positions of the heat sources, $T(\vec{r} \to S) = T_0$, where $T_0$ = 36.6 °C is the normal temperature of a human body.

*Spherically symmetric model*

First, let us consider the simplest case of a spherically symmetric distribution of heat sources

$$n(r) = \begin{cases} q_0 C_m, & r \leq R_m \\ 0, & r > R_m \end{cases} ; \quad R_m < R_0 \quad (2)$$

where $q_0$ is the SAR of the assembly (W/g), $C_m$ is the concentration of magnetic nanoparticles (g/cm$^3$) uniformly distributed in a spherical region of radius $R_m$, $R_0$ is the outer radius of the domain under consideration. Under the assumption that the temperature distribution approaches a constant value at the domain boundary, $T(R_0) = T_0$, the solution of Eqs. (1), (2) for the temperature increment $\Delta T = T(r) - T_0$ has the form [14-17]

$$\Delta T(r) = \begin{cases} \frac{q_0 C_m}{6 k_s}(R_m^2 - r^2) + \frac{q_0 C_m R_m^2}{3 k_s}\left(1 - \frac{R_m}{R_0}\right), & r < R_m \\ \frac{q_0 C_m R_m^3}{3 k_s}\left(\frac{1}{r} - \frac{1}{R_0}\right), & R_m < r < R_0 \end{cases} \quad (3)$$

Using Eq. (3) it is easy to show that in the case $R_m \ll R_0$ the temperature distribution only weakly depends on the external radius $R_0$ of the domain under consideration.

The SAR of the assembly is determined by the magnetic and geometric properties of the nanoparticles [2-4]. At a fixed frequency and amplitude of ac magnetic field parameter $q_0$ can be considered given. The average thermal conductivity of the biological medium is known to be $k_s$ = 0.49 W(mK)$^{-1}$ [37]. Thus, only the radius of the heated region $R_m$ and the concentration of magnetic nanoparticles $C_m$ distributed in this region are the variables of the spherically symmetric model, Eq. (3). Moreover, it is easy to see that in the reduced coordinates, $x = r/R_m$, in the limit $R_m \ll R_0$ the temperature distribution in the region under consideration depends only on a single parameter, $q_0 C_m R_m^2 / k_s$.

Let us fix the typical values $q_0$ = 100 W/g and $k_s$ = 0.49 W(mK)$^{-1}$, and consider the dependence of the temperature distribution on the parameters $R_m$ and $C_m$ that can be selected by a researcher.

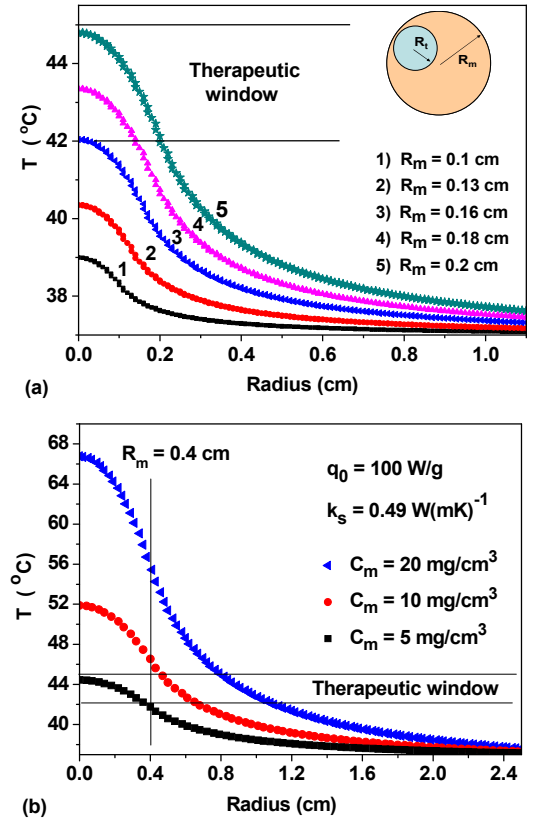

Figure 1. a) The temperature distribution in the spherically symmetric model, Eq. (3), for fixed values $q_0$ = 100 W/g, $k_s$ = 0.49 W(mK)$^{-1}$, and $C_m$ = 20 mg/cm$^3$ depending on the radius of the heating region $R_m$; b) the same for $R_m$ = 0.4 cm for various concentration of nanoparticles $C_m$ in the heating region.

In magnetic hyperthermia typical values of the nanoparticle concentration are given by $C_m$ = 5 – 30 mg/cm$^3$, whereas the characteristic radii of tumors in laboratory animals are within the range $R_t$ = 0.1 – 0.8 cm [4-9].

Let us consider first the case of a relatively small heating area with radius $R_m \leq 0.2$ cm. As Fig. 1a shows, even at concentration $C_m$ = 20 mg/cm$^3$ the therapeutic window, 42 < T < 45° C, appears only for $R_m \geq 0.16$ cm. Moreover, at $R_m$ = 0.16 cm the heating of only very small tumors is possible. On the other hand, at $R_m$ = 0.2 cm any tumor with $R_t$ < 0.2 cm can be heated if it is completely placed within the heated region, as the inset in Fig. 1a shows. Thus, small isolated tumors with $R_t \leq 0.1$ cm can be confidently warmed up, covering them with heating areas of radius $R_m$ = 0.2 cm. At the same time, according to Fig. 1a, a significant surrounding spherical layer of healthy tissues in the range of radii 0.2 < r < 0.8 cm also warms up in the temperature range 38 - 42° C.

Consider now the case of a relatively large heating region $R_m$ = 0.4 cm for various values of $C_m \leq 20$ mg/cm$^3$. As Fig. 1b shows, a huge overheating, T > 50 °C, occurs in the central part of the heated region at concentrations $C_m \geq 10$ mg/cm$^3$. This heating mode is actually a thermal ablation. It should be excluded in stationary conditions. Moreover, even an increase in the particle concentration from $C_m$ = 5 mg/cm$^3$ to $C_m$ = 10 mg/cm$^3$ is already quite critical, since it leads to elevated temperatures in the central part of the heated region. Therefore, for heating region $R_m$ = 0.4 cm the concentrations of the order of $C_m$ = 5 - 6 mg/cm$^3$ seem suitable. But at the same time, only the heating of tumors with $R_t < R_m$ is possible if the heated area completely cover the tumor. It is clear that this condition is restrictive if the tumor has a complex shape, since in this case the temperature in some areas of the tumor may not fall into the therapeutic window.

This illustrative example shows that one of the problems in magnetic hyperthermia is the formation of an optimal temperature distribution for the treatment of tumors of irregular shape or sufficiently large sizes, $R_t$ > 0.4 - 0.5 cm.

*Linear combination of heat sources*

It has been shown in the previous section that the spherically symmetric model, Eq. (3), is not sufficiently flexible for use in planning the temperature distribution in magnetic hyperthermia. However, there is an interesting possibility of covering a large tumor of complex shape with thermal fields from several quasi-spherical heat sources.

It is easy to see that due to the linearity of Eq. (1), in the presence of several spherical heating regions, $n_i(|\vec{r} - \vec{\rho}_i|)$, i = 1, 2, …, N, whose centers are located at the points $\vec{\rho}_i$, the solution of Eq. (1) is a linear combination of solutions (3) of the form

$$\Delta T(x,y,z) = \sum_i \Delta T_i(|\vec{r} - \vec{\rho}_i|).$$

Moreover, if the external boundary of region S is located far enough from the domain of heat sources, then the boundary condition, $\Delta T(x,y,z)|_S = 0$, is approximately satisfied too.

In the presence of several heat sources, a sufficient number of fitting parameters appear in the problem, such as the centers $\vec{\rho}_i$ of location of heat sources, their radii $R_{m,i}$, as well as the concentration of magnetic nanoparticles in various sources, $C_{m,i}$. As a result, it becomes possible to design an optimal stationary temperature distribution inside and around a tumor of a complex shape.

As an example, consider a tumor of irregular shape located near the center of a cube with an edge of 1 cm. Fig. 2a shows the temperature distribution in the central cross section of the cube, z = 0.5 cm, created by a single spherical heat source located in its center. The parameters of the heat source are given by $q_0$ = 150 W/g, $C_m$ = 20 mg/cm$^3$ and $R_m$ = 0.15 cm, correspondingly. The thermal conductivity of the medium $k_s$ = 0.49 W(mK)$^{-1}$. As Fig. 2a shows, in the presence of a single heat source with the given parameters the edges of the tumor do not fall into the therapeutic interval. This may lead to a recession of the tumor after some time from the end of the therapeutic procedure. On the other hand, as Fig. 2b shows in the presence of four similar heat sources located in the plane z = 0.5 cm, the tumor warms up completely. However, in this case in its central part the temperature is significantly higher than the therapeutic window, which leads to a thermal ablation in this area.

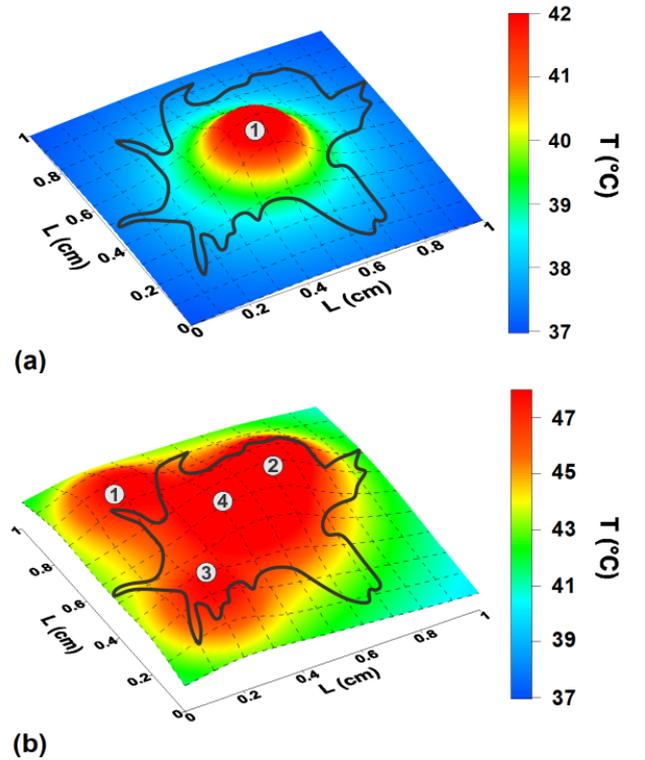

Figure 2. a) The spatial temperature distribution in the central cross section, z = 0.5 cm, of a cube with size 1 cm created by a single heat source in the cube center; b) the same for the case of four identical heat sources with centers lying in the plane z = 0.5 cm and marked with the points 1 - 4. The contours of an irregularly shape tumor in the given cross sections are shown by a black curve.

Fortunately, the calculations show that satisfactory results on modeling the thermal field in a tumor can be obtained with a appropriate distribution of heat sources of the same thermal power. In this case, the number of fitting parameters in the optimization problem decreases significantly. Namely, it is sufficient to select a spatial period of the heat source location and a thermal power of an individual source.

As an example, Fig. 3 shows the various cross-sections of the temperature distribution in a cube with an edge of 1 cm created by six identical heat sources located symmetrically in the cube at a distance of 0.3 cm from the cube center. The thermal parameters of each source are $q_0$ = 100 W/g, $C_m$ = 15 mg/cm$^3$ and $R_m$ = 0.15 cm, respectively. As Fig. 3 shows, this choice of heat sources creates a nearly uniform temperature distribution near cube center with a temperature increment $\Delta T$ = 5 - 6 °C, so that almost the entire heating domain enters the therapeutic window. Apparently, the suitable distribution of identical heat sources is preferable for heating a tumor of large size or complex geometry located in a homogeneous biological environment. To provide necessary spatial distribution of heat sources in a tumor volume promising methods of minimally invasive and regenerative therapeutics [38] can be used.

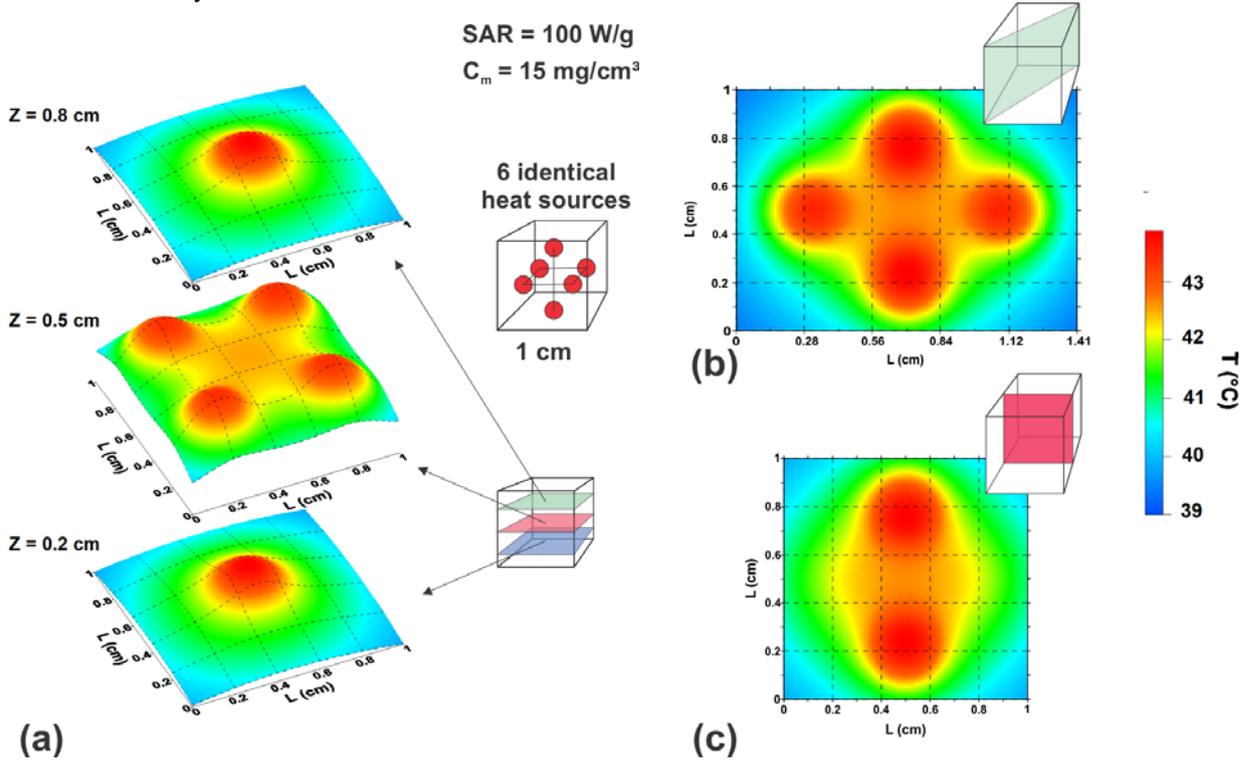

Figure 3. The spatial temperature distribution in biological media created by 6 heat sources of the same power symmetrically located with respect to cube center. a) cross – sections of the cube by the planes $z/L_z$ = 0.2, 0.5 and 0.8, respectively; b), c) the same for the planes $(x+y)/L_z$ = 1 and $x/L_z$ = 0.5, correspondingly.

It is interesting to note that a moderate SAR value, $q_0$ = 100 W/g, turns out to be sufficient to provide optimal temperature distribution in a relatively large domain of biological medium with a diameter $D \sim 0.5 - 0.8$ cm. It is worth noting that the temperature distribution, Eq. (3), generated by a single heat source depends on the product $q_0 C_m$. For the temperature distribution shown in Fig. 3 this product is given by 1.5 W/cm$^3$. Furthermore, the total amount of magnetic nanoparticles introduced into the heating volume is estimated to be only $m = 6 C_m V_m \approx$ 1.3 mg, where $V_m = 4\pi R_m^3/3$ is the volume of the single heat source.

**B. Optimization of the heat sources**

We have already noted in the introduction the difficulties that arise in magnetic hyperthermia if one uses assemblies consisting of individual superparamagnetic nanoparticles. Actually, in a biological environment an uncontrolled agglomeration of nanoparticles usually occurs. In addition, nanoparticles often remove from a tumor with blood flows. A serious problem is a rapid destruction of individual nanoparticles in the body under the influence of an aggressive media, etc. It seems that at least some of these problems can be overcome if small optimized capsules of magnetic nanoparticles are used as heat sources. In this case the particles will be better protected from the aggressive influence of the biological media. Besides, the effect of MD interaction on SAR could be minimized with a proper capsule design. In the literature [33 - 36] there are some successful examples of the experimental creation of small capsules of magnetic nanoparticles. However, the optimization of the capsules structure to make them most suitable for magnetic hyperthermia has not yet been discussed.

There are arguments [33] showing that an average diameter of magnetic capsules should not exceed $D_c$ = 100 - 150 nm, since capsules of this size, covered with thin biocompatible envelopes, easily penetrate tumors. The next in importance is the question of the magnetic and geometric parameters of the nanoparticles themselves. Although magnetic iron oxides are currently the most popular in biomedicine [4-6,23-25], it is clear that in magnetic hyperthermia the use of metallic iron nanoparticles [39-41] would be preferable due to the much higher saturation magnetization of iron. Regardless of the type of ferromagnet used, to create magnetic capsules it is necessary to use particles of optimal diameters taking into account that the window of optimal diameters depends also on the frequency and amplitude of applied ac magnetic field [30,31]. Finally, it is worth combining into capsules nanoparticles previously coated with rather thick non-magnetic shells, comparable to the particle diameter. This significantly reduces the intensity of MD interaction between the particles inside the capsule [31] and improves the protection of nanoparticles from the aggressive action of the media.

Based on these considerations, we have optimized the structure of small capsules of iron nanoparticles to be used in magnetic hyperthermia. The details of the numerical simulation are given in Refs. [30,31]. The optimization procedure is carried out at moderate values of the amplitude and frequency of ac magnetic field, which is preferable for the application of this technique in clinical practice [10,11]. To be specific, we consider two characteristic cases for ac magnetic field parameters: 1) $H_0$ = 50 Oe, $f$ = 200 kHz and 2) $H_0$ = 100 Oe, $f$ = 100 kHz.

Fig. 4a shows the results of SAR calculations for a non-interacting randomly oriented assembly of spherical iron nanoparticles with cubic type of magnetic anisotropy, depending on the particle diameter for $f$ and $H_0$ values specified above. The saturation magnetization and cubic magnetic anisotropy constant of the particles are given by $M_s$ = 1700 emu/cm$^3$ and $K_c$ = 4.6×10$^5$ erg/cm$^3$, respectively [42]. As Fig. 4a shows, significant absorption of the energy of ac magnetic field occurs in rather narrow intervals of particle diameters. For the first case, $H_0$ = 50 Oe, $f$ = 200 kHz, the optimal window of particle diameters is $D$ = 18 - 20 nm, where SAR reaches values of the order of 350 - 430 W/g. For the second case, $H_0$ = 100 Oe, $f$ = 100 kHz, the SAR exceeds 500 W/g in the range of diameters $D$ = 21 - 23 nm.

Of course, SAR of an assembly of capsules usually decreases [35] due to the influence of strong MD interaction of closely spaced nanoparticles inside the capsules. To minimize this effect, it was proposed [31] to cover particles with nonmagnetic shells with a thickness $t_{Sh}$ of the order of the particle diameter, $t_{Sh} \sim D$. The optimization problem consists, therefore, in randomly placing within a capsule a number of nanoparticles of optimal diameters $D$, covered with nonmagnetic shells of thickness $t_{Sh}$, under the condition that the diameter of the capsule $D_c$ does not exceed 100 – 150 nm.

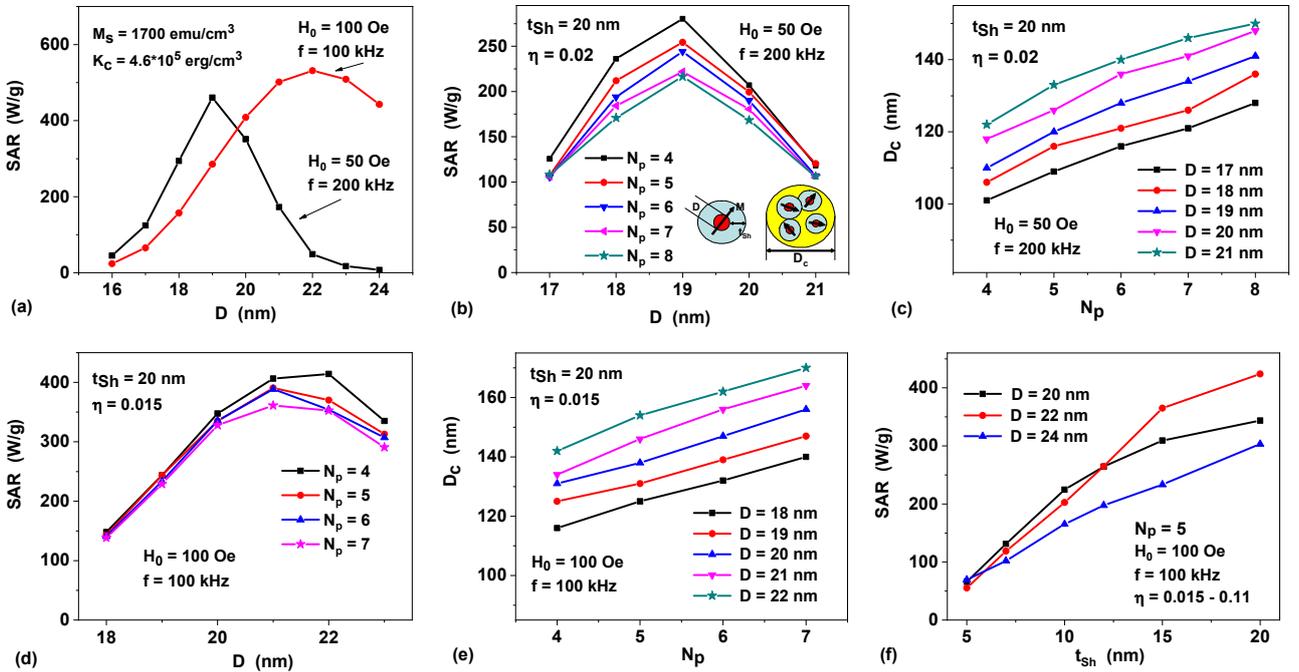

Figure 4. a) SAR of an assembly of non interacting randomly oriented spherical iron nanoparticles as a function of particle diameter for various frequencies and ac magnetic field amplitudes; b) SAR of a dilute assembly of magnetic capsules as a function of particle diameter for different number of nanoparticles within the capsule; c) the size of the capsules $D_c$ as a function of particle number $N_p$ for different optimal particle diameters. For the panels b) and c) $H_0$ = 50 Oe, $f$ = 200 kHz, $\eta$ = 0.02. Inset in Fig. 4b shows the geometry and arrangement of nanoparticles into a capsule. d), e) the same as for the panels b) and c), but for the case $H_0$ = 100 Oe, $f$ = 100 kHz, $\eta$ = 0.015; f) SAR of a dilute assembly of capsules as a function of non-magnetic shell thickness $t_{Sh}$ for various optimal particle diameters.

Fig. 4b shows the results of SAR calculation for dilute assemblies of magnetic capsules for the case $H_0$ = 50 Oe, $f$ = 200 kHz. For iron nanoparticles with diameters in the range $D$ = 17 - 21 nm, covered with nonmagnetic shells of fixed thickness $t_{Sh}$ = 20 nm, SAR values are shown for various numbers $N_p$ = 4 - 8 of nanoparticles randomly distributed in a capsule volume. For each set of $D$ and $N_p$ values in Fig. 4b the calculation results are averaged over assemblies of 400 independent realizations of individual capsules. The volume fraction of magnetic material in a capsule is defined as $\eta = N_p V/V_c$, where $V = \pi D^3/6$ is the volume of a particle and $V_c = \pi D_c^3/6$ is the volume of a capsule, respectively. In the calculations performed the volume fraction was maintained approximately constant $\eta \approx$ 0.02, which is close to its maximal possible value for particles covered with the non-magnetic shells of the given thickness $t_{Sh}$ = 20 nm. Evidently, in random capsules constructed in this way, the distance between the centers of magnetic nuclei cannot be less than $d_{sep}$ = $D + 2t_{Sh}$. This leads to a significant decrease in the characteristic energy of MD interaction in capsule volume. It is interesting to note that the maximum of SAR in Fig. 4b corresponds to the particle diameter $D$ = 19 nm, as for the assembly of non interacting nanoparticles. But the SAR value at the maximum is only 220-270 W/g depending on the number of particles in the capsule. Though SAR decreases in comparison with the corresponding value of 430 W/g for the assembly of non-interacting nanoparticles, it is still sufficient for application in magnetic hyperthermia. It is important to note also that as Fig. 4c shows, the capsule diameters in all cases considered fall within the range $D_c$ = 100 - 150 nm.

Fig. 4d shows SAR of a dilute assembly of capsules calculated for the case $H_0$ = 100 Oe, $f$ = 100 kHz, $\eta \approx$ 0.015. With an increase in the ac magnetic field amplitude, the maximum SAR value for the capsule assembly shifts to the range $D$ = 21-22 nm. It practically coincides with the position of the SAR maximum for non-interacting iron nanoparticles at the same values of $H_0$ and $f$. However, the SAR value at the maximum for the assembly of capsules reduces to the value 350-400 W/g. Fig. 4e shows the diameters of the magnetic capsules $D_c$ versus the diameter and the number of particles in the capsule, respectively. For optimal particle diameters $D$ = 21-22 nm the capsule diameter turns out to be in the range $D_c$ = 140 - 170 nm if the number of particles in the capsule $N_p \leq 7$.

Fig. 4f shows the results of a similar optimization procedure for particle various diameters, $D$ = 20 - 24 nm, for the case when the number of particles in the capsule is fixed, $N_p$ = 5, but the thickness of the nonmagnetic shell of the particles varies. In the calculations performed the minimum capsule diameter $D_c$ is determined under the condition that $N_p$ = 5 particles of diameter $D$, covered with shells of thickness $t_{Sh}$, can be arranged in the capsule volume. As $t_{Sh}$ decreases from 20 to 5 nm, the minimum capsule diameter $D_c$ decreases from 180 to 80 nm. Simultaneously, the volume fraction of magnetic material increases from $\eta$ = 0.015 to $\eta$ = 0.11 and the minimum distance $d_{sep}$ between the centers of magnetic nuclei decreases accordingly. Thus, Fig. 4f confirms that a decrease in the particle shell thickness leads to a significant decrease in the SAR of the dilute assembly of magnetic capsules due to increase in the intensity of the MD interaction.

Let us estimate the heating ability of magnetic capsules designed using metallic iron nanoparticles. As we mentioned above, to generate typical thermal distribution in biological media shown in Fig. 3 it is sufficient to provide the heating power of a single heat source at a level of 1.5 W/cm$^3$. The SAR calculations show that the optimized magnetic capsules can give much higher heating power in a capsule volume. Actually, as Fig. 4d shows, for assembly of capsules containing $N_p$ = 4 nanoparticles with diameter $D$ = 22 nm the SAR maximum is given by 414 W/g. Taking into account the average capsule diameter $D_c$ = 142 nm and iron density $\rho$ = 7.8 g/cm$^3$ it is easy to calculate that this SAR value corresponds to average heating power around 48 W/cm$^3$ in the capsule. Suppose now that such magnetic capsules are uniformly distributed in the volume of a single heat source. To maintain the heating power of the whole source at the level 1.5 W/cm$^3$ a balance relation $48V_c = 1.5L^3$ has to be satisfied, where $L$ is the size of a cube containing one capsule in average. From this relation one obtains approximately $L \approx 2.5D_c$. Evidently, the size $L$ is an average distance between the centers of the magnetic capsules distributed in the heat source volume. This distance seems sufficiently large to reduce considerably the MD interaction between the capsules. Next, in the case considered in the volume $L^3$ there are only 4 iron nanoparticles in average. Therefore, the iron nanoparticle concentration in the heat source is estimated to be as low as $C_m = \rho N_p V/L^3 \approx$ 3 mg/cm$^3$.

It is interesting to note that in the remarkable paper [35] magnetic nanoparticle-loaded polymer nanospheres with an average diameter $D_c$ = 80 - 100 nm were successfully synthesized. They contain mono-dispersed MnFe$_2$O$_4$ particles with diameters $D$ = 6 and 18 nm, respectively. The polymer nanospheres exhibit excellent colloidal stability in both water and physiological environment. For individual nanoparticles and assemblies of nanospheres with various particle density SAR measurements were carried out in water and agarose gel in ac magnetic field with $H_0$ = 50 Oe and $f$ = 435 kHz. It was shown that the SAR of individual particles in water substantially depends on the particle diameter. Namely, for particles with diameters $D$ = 6 and 18 nm, SAR values of 83 and 580 W/g were obtained, respectively. This result shows once again the importance of determining the optimum particle diameter when the SAR reaches its maximum. Unfortunately, the authors [35] limited themselves to studying particles of two diameters only.

Important results were also obtained [35] for assemblies of nanospheres containing particles with diameter $D$ = 18 nm with different particle loading ratio. For nanosphere assembly with a lower loading ratio SAR value of 332 W/g was obtained, which is three times higher than for nanospheres of the same diameter with a higher loading ratio. This result directly shows that the SAR of the assembly of nanospheres decreases with an increase in the intensity of MD interaction. Finally, it was found that in contrast to the assembly of individual nanoparticles, for both types of assemblies of nanospheres with low and high loading ratio the SAR of the assembly in water and agarose gel is nearly the same. This means that the Brownian contribution to heat generation in the nanospheres under the influence of ac magnetic field can be neglected.

Note, in this regard, that the SAR calculations shown in Fig. 4 were carried out under the assumption that the rotation of magnetic nanoparticles or capsules in the surrounding media is inhibited.

## 3. Conclusions

An important issue in planning magnetic hyperthermia is the control of the temperature distribution in a tumor and healthy tissues surrounding it. During a successful medical procedure it is necessary to maintain a temperature in the range of 42 - 45°C in the entire tumor volume, which may have a rather complex shape. Furthermore, the heating of healthy tissues surrounding the tumor above 42°C should be avoided for the duration of the hyperthermia session that is about 20-30 minutes.

In this paper it is proposed to use a linear combination of spherically symmetric heat sources when planning the optimal temperature field in magnetic hyperthermia. This is especially important if the tumor has a large size or complex shape. In this approach a sufficiently large number of fitting parameters appears in the problem, namely, the number of sources, the position of their centers, the thermal power and sizes of individual sources, etc. In this paper it is shown that using suitable arrangement of identical heat sources in some cases it is possible to obtain a temperature distribution close to optimal. We have also shown that with an optimal distribution of heat sources in the biological environment, heat sources with moderate SAR values of the order of 100 - 150 W/g can be used to perform successful medical thermal treatment.

Further, important problems in magnetic hyperthermia are the agglomeration of magnetic nanoparticles in tissue, the removal of the nanoparticles from a tumor by means of a blood flow, and the rapid destruction of magnetic nanoparticles in an aggressive biological environment. All these processes lead to a significant decrease of the heating power of magnetic nanoparticles in the biological medium over time and make the heat effect on the tumor inefficient. In a number of works [33 - 36] it was proposed to use in magnetic hyperthermia small capsules containing magnetic nanoparticles as local heat sources. The biocompatible material of the capsule will reliably protect magnetic nanoparticles from destruction. Therefore, such capsules will be able to maintain their heating power in a tumor for a sufficiently long time. In the present paper a design of magnetic capsules promising for application in magnetic hyperthermia is developed. It is suggested to cover the nanoparticles to be combined into a capsule by non magnetic shells of sufficient thickness to reduce the intensity of MD interaction between the particles inside the capsule. The diameters of nanoparticles used to create magnetic capsules, and the number of nanoparticles in the capsule volume were optimized using numerical simulation.

## Acknowledgments

Authors gratefully acknowledge the financial support of the Ministry of Science and Higher Education of the Russian Federation in the framework of Increase Competitiveness Program of NUST «MISIS», contract № K2-2019-012.

___